\documentclass[12pt,english,aps]{revtex4}
\usepackage{babel}
\usepackage{amsmath}
\usepackage{amssymb}
\usepackage{graphicx}
\usepackage[unicode=true,pdfusetitle,
 bookmarks=true,bookmarksnumbered=false,bookmarksopen=false,
 breaklinks=false,pdfborder={0 0 1},backref=false,colorlinks=false]
 {hyperref}
\usepackage{breakurl}

\makeatletter
\@ifundefined{textcolor}{}
{%
 \definecolor{BLACK}{gray}{0}
 \definecolor{WHITE}{gray}{1}
 \definecolor{RED}{rgb}{1,0,0}
 \definecolor{GREEN}{rgb}{0,1,0}
 \definecolor{BLUE}{rgb}{0,0,1}
 \definecolor{CYAN}{cmyk}{1,0,0,0}
 \definecolor{MAGENTA}{cmyk}{0,1,0,0}
 \definecolor{YELLOW}{cmyk}{0,0,1,0}
}

\usepackage{epsfig}\usepackage{color}\usepackage{amsfonts}\usepackage{amsthm}\usepackage{comment}
\newcommand{\ltwid}{\mathrel{\raise.3ex\hbox{$<$\kern-.75em\lower1ex\hbox{$\sim$}}}}
\newcommand{\gtwid}{\mathrel{\raise.3ex\hbox{$>$\kern-.75em\lower1ex\hbox{$\sim$}}}}

\newcommand{\bea}{\begin{eqnarray}}
\newcommand{\eea}{\end{eqnarray}}
\newcommand{\Tr}{{\rm Tr}}
\theoremstyle{remark}

\renewcommand\[{\begin{equation}}
\renewcommand\]{\end{equation}} 

\makeatother

\begin{document}

\title{Closed hierarchies and non-equilibrium steady states of driven systems}

\author{Israel Klich}

\affiliation{Department of Physics, University of Virginia, Charlottesville, 22904, VA,
USA}
\begin{abstract}
We present a class of tractable non-equilibrium dynamical quantum systems which includes combinations of injection, detection and extraction of particles interspersed by unitary evolution. We show how such operations generate a hierarchy of equations tying lower correlation functions with higher order ones. The hierarchy closes for particular choices of measurements and leads to a rich class of evolutions whose long time behavior can be simulated efficiently. In particular, we use the method to describe the dynamics of current generation through a generalized quantum exclusion process, and exhibit an explicit formula for the long time energy distribution in the limit of weak driving. 
\end{abstract}
\maketitle
\tableofcontents
\section{Introduction}
Significant activity has been devoted to the study of quantum systems out of equilibrium, with a rapid increase in interest due to the relevance to experiments with ultra-cold atomic gases, whose coherent evolution may be effectively controlled and decoupled from dissipation to a heat bath \cite{bloch2008many,polkovnikov2011colloquium,lamacraft2012potential}.
Non equilibrium dynamics is typically studied in processes such as external driving, repeated quantum measurements and quantum quenches.
The fundamental question that arises in such cases is what is the
long term behavior of the system: does it eventually reach a non-equilibrium steady
state? What is the nature of such a state?

In studying the aforementioned non-equilibrium situations, some highly successful tools of equilibrium statistical physics, such as linear response theory, may easily fail.  Thus, there is a need to develop new methods to deal with some of these problems. Here we focus on one  such idea - that of establishing closed hierarchies in order to get tractable equations for correlation functions. 
Specifically, in many statistical mechanics problems, it is possible to make a
systematic connection between the evolution of $n$ body density functions
with $n+1$ density functions. A prime example for
such a set of relations is the Bogoliubov-Born-Green-Kirkwood-Yvon
(BBKGY) hierarchy, which is the essential structure leading to the
Boltzmann equation. In the Boltzmann equation, single particle densities are tied to higher order
correlation functions represented in the collision integral (see, e.g. \cite{bonitz1998quantum}). In this letter, we describe the requirements on obtaining a hierarchy under general quantum operations on fermions. We then show how the hierarchy may be closed for a quantum system that is periodically evolved, detected, and injected with current. Finally, we use the idea to describe dynamics of current buildup, and the energy distribution in the long term non-equilibrium steady state.

To begin the discussion, consider the most general evolution of a density matrix, describing
unitary evolution, measurements and interaction
with the environment. Written as
\begin{eqnarray}
\rho\to{\cal L}(\rho)=\Sigma_{\nu}A_{\nu}\rho A_{\nu}^{\dagger}\text{ };\text{ }\Sigma_{\nu}A_{\nu}^{\dagger}A_{\nu}=1\label{eq:General Evolution}
\end{eqnarray}
This form ensures $\rho$ remains a non-negative matrix, and the normalization
condition on the Krauss operators $A_{i}$ ensures
that $\Tr\rho=1$ is preserved under the evolution.

In general, there is no simple relation between correlation functions
computed in  state $\rho$ before and after the evolution (\ref{eq:General Evolution}), which necessitates working in an exponentially large Hilbert space and is therefore often un-tractable. 

Hierarchy structures have been used before in the context of Kossakowski-Lindblad evolution, which is a particular limit of \eqref{eq:General Evolution}. For example, the steady state of a dissipative XX spin chain in the presence of driving and dissipation has been studied extensively \cite{temme2012stochastic,vznidarivc2010exact,vznidarivc2011solvable,eisler2011crossover}. 
Also, conditions for a closed hierarchy in the continuous time frame work where also stated in \cite{vzunkovivc2014closed,caspar2016dynamics,mesterhazy2017solvable}. Here we concentrate on a discrete time framework, but also supply corresponding Kossakowski-Lindblad results as a special limit. In other processes, the possibility of getting a closed equation for Kossakowski-Lindblad evolution of noise averaged expectation values was studied in \cite{rahmani2014dynamics}, to explore the stability of fractional charges to noisy hopping processes.

\begin{figure}
 \centering (a) \includegraphics[width = 0.4\textwidth]{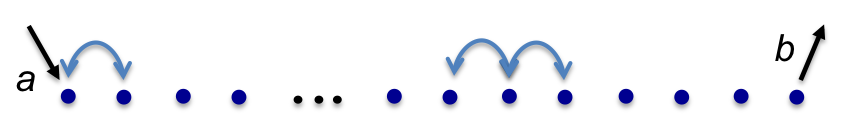} \vskip -0.1cm 
(b) \includegraphics[width = 0.4\textwidth]{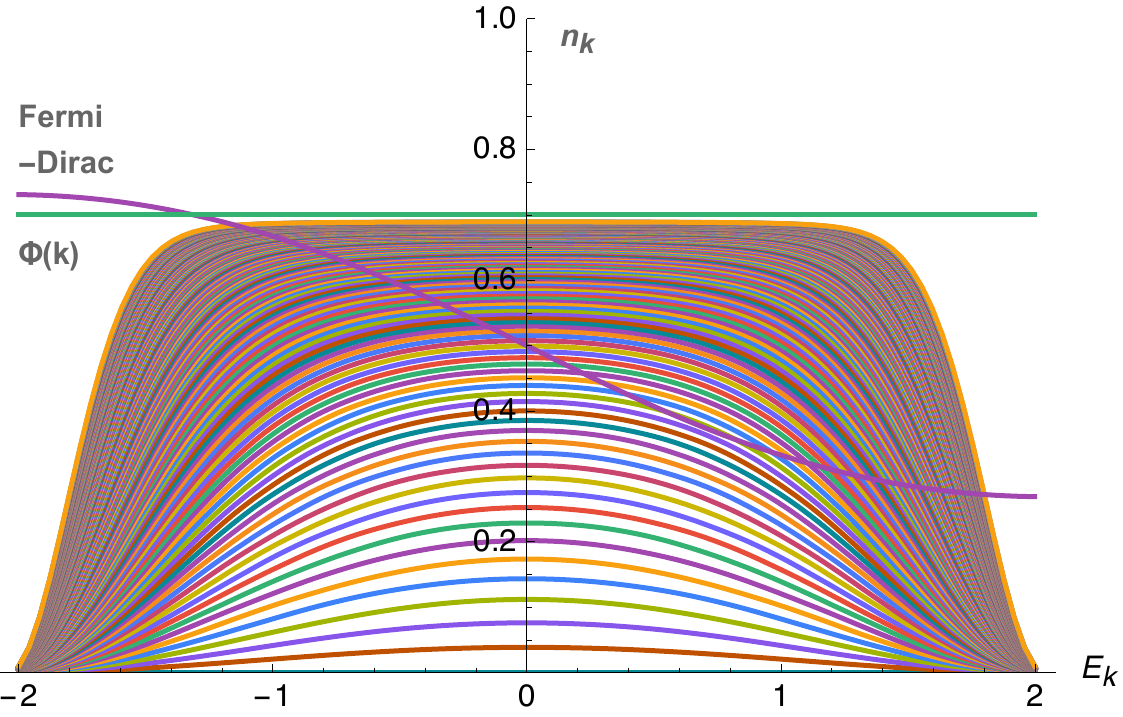}  \vskip 0.1cm
 $~~~~~$\includegraphics[width = 0.4\textwidth]{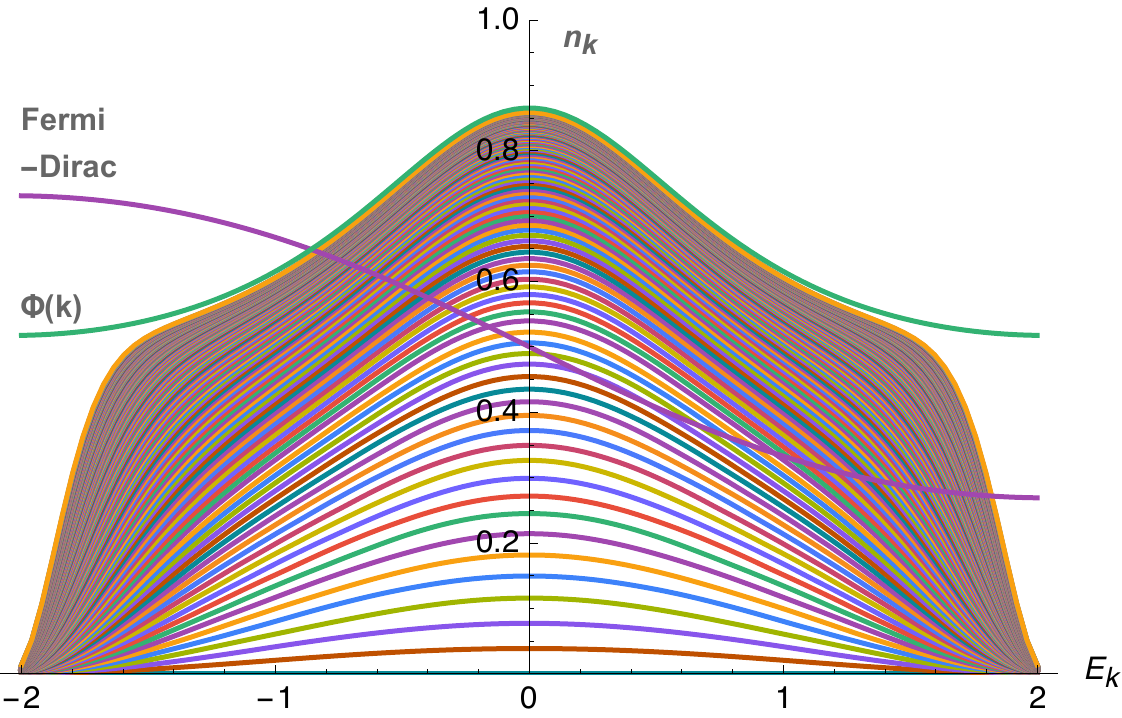}  
\caption{(a) The fermion hopping model.  (b) Approach to $\Phi(k)$ for $ {r} =0.01, {\alpha} =0.7,\tau=0.1$. Results for extraction at $b=100$ (upper panel) and $b=99$ (lower panel). $300$ iterations between successive curves. For reference a Fermi-Dirac distribution is shown. }\label{steadycurrent}
\end{figure}

We utilize the power of this approach to study a non-equilibirum process of current generation, as schematically depicted in Fig. \ref{steadycurrent} (a).
In this process, we connect site $a$ to a lead, where a current is injected, and particles are allowed to go out at site $b$ (two choices for $b$ are shown).  The process is explicitly described by
\bea \label{current injection process} &
\rho \longrightarrow  U  ((1- {r} )\rho
   + {r}   {\alpha}   [\epsilon_a(2-\epsilon_a) a_a^{\dagger }\rho  a_a+(1-\epsilon_a(1-n_a))\rho 
   (1-\epsilon_a(1-n_a)) ]+\\ & \nonumber  {r} (1- {\alpha} ) [\epsilon_b(2-\epsilon_b)a_b \rho  a_b^{\dagger
   }+ (1-\epsilon_b n_b )\rho 
    (1-\epsilon_b n_b ) ] )U^{\dagger },
   \eea
where $n_{a/b}=a_{a/b}^{\dagger}a_{a/b}$ checks for the presence of a fermion on the injection/extraction site, and $U=e^{-i \tau \sum h_{nm}a^{\dag}_na_m}$ describes evolution between attempts during a time interval $\tau$.  Here $ {r} $ is the overall attempt rate, $ {\alpha} $ is the relative probability of injecting vs extracting attempts, and $\epsilon_{a,b}$ are related to the efficiency of the injection/extraction attempts: when $\epsilon_{a,b}=1$, particle injection or removal happens with probability $1$ if an attempt is made.
We show below that this process leads to a closed equation \eqref{G evolution} for the two point function of the system, which can be then computed numerically. 
It is important to emphasize that the long time steady state reached by the system is not a thermal equilibrium state, in that the energy occupation is very different from a Fermi-Dirac distribution governed by the single particle Hamiltonian $h$ governing the evolution $U$. 

For small $ {r} $, we find a remarkable asymptotic formula for the steady state distribution $\Phi_k\equiv \langle steady| a_k^{\dag}a_k| steady\rangle$. Here $k$ labels the eigenstates $|k \rangle$ of the single particle hamiltonian $h_{nm}$, $h{|}k \rangle=E_k|k \rangle$. Let $p_{a,k}=|\langle a|k\rangle|^2, p_{b,k}=|\langle a|k \rangle|^2$ be overlaps of these states with the sites $a,b$. Then $\Phi_k$ is a function of the ratio $p_{a,k}/p_{b,k}$:
\bea\label{energy distribution}
\Phi_k= {{\cal A}+{\cal B} ~{p_{a,k}\over p_{b,k}} \over (1-\alpha)\epsilon_b+\alpha \epsilon_a {p_{a,k}\over p_{b,k}}} \label{main_distribution}
\eea
Note the appearance of the relative injection rates/extraction rates: $\alpha \epsilon_a$ and $(1-\alpha)\epsilon_b$.

The coefficients ${\cal A},{\cal B}$ are given below in Eq. \eqref{A and B coefficients}. We emphasize that this expression is valid for any system obeying the form \eqref{current injection process}, and is non perturbative.

In the limit of low tunneling probability, $\epsilon_a,\epsilon_b\rightarrow 0$, the result depends only on the ration of injection to removal rates and simplifies to:
\bea \label{low_tunneling_limit}
\Phi _k\sim \frac{\frac{\alpha  \epsilon_a}{(1-\alpha )\epsilon_b}\text{  }\frac{
   p_{a,k}}{p_{b,k}}}{1+\frac{\alpha  \epsilon_a}{(1-\alpha )\epsilon_b}
   \frac{ p_{a,k}}{p_{b,k}}}
\eea
This last expression has a simple interpretation: the probability of occupying a given mode $k$ is determined by the ratio between the effective tunneling probability into energy $k$ from site $a$ compared to the effective tunneling rate of the state $k$ through site $b$.
The limit of $r,\epsilon_a,\epsilon_b\rightarrow 0$, also corresponds to the limit where a Kossakowski-Lindblad equation can be used to describe \eqref{current injection process}. Indeed, as we show below, one can obtain \eqref{low_tunneling_limit} from Kossakowski-Lindblad treatment of the process \eqref{current injection process}.  

We stress that in the low tunneling limit, the steady state $\Phi_k$ does not depend on system details except the tunneling rates and the probabilities $p_{a/b,k}$. However, 
going back to the formula \eqref{energy distribution}, the details of the distribution depend of sensitively on the choice of parameters. In particular, we note that even if $p_{a,k}=0$, i.e. there is no overlap between a given energy mode and the insertion site (or mode), $\Phi_k$ can be non vanishing, due to higher order processes, a feature which is absent in the simpler Kossakowski-Lindblad limit expression \eqref{low_tunneling_limit}. This feature illustrates the non-perturbative dependence of $\Phi_k$ on the system parameters (and on $\epsilon_a,\epsilon_b,\alpha$). 

\begin{figure}
 \centering  \includegraphics[width = 0.4\textwidth,trim={0 0 0 0.05cm},clip]{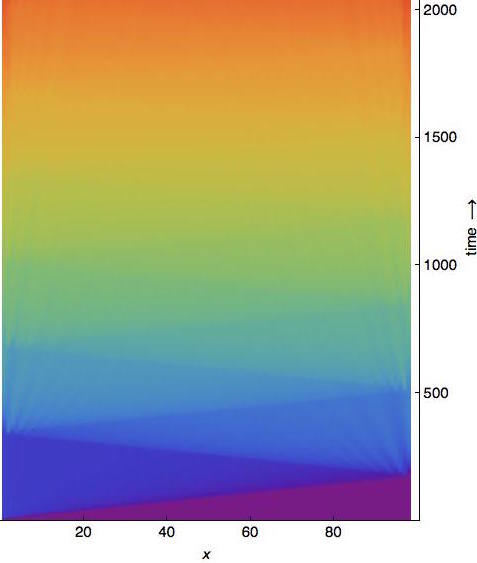}  \vskip -0.1cm
\caption{The fermion hopping model: Evolution of local density, $\langle a_i^{\dag}a_i\rangle$, in space and time (red/blue corresponds to high/low density, same parameters). }\label{steadycurrent1}
\end{figure}

For illustration, we consider hopping on a chain of length $N$, with the standard Hamiltonian $H_{hop}=\sum_{i=1}^{N-1} a_i^{\dag}a_{i+1}+h.c.$ corresponding to Dirichlet boundary conditions. In this case  $p_{a,k}/p_{b,k}=\sin^2({\pi a k\over N+1})/\sin^2({\pi b k\over N+1})$. In Fig. \ref{steadycurrent} we illustrate the result with $N=100$, and injection at $a=1$. We evolve the system from an initial vacuum state at $t=0$. The results for extraction at the final and penultimate sites $b=100,99$ respectively, show sensitivity to the choice of operation sites. The energy distribution is computed numerically at long times and is clearly seen to approach $\Phi$ in the long time limit. 
We stress that  once driving has stopped, the energy distribution  $\Phi$ will remain the stationary distribution under the subsequent free evolution. Fig \ref{steadycurrent1} shows the actual evolution of the density as we inject the particles into the system.

\section{General framework}
We now turn to establishing the framework for our processes. We consider a system of fermions on a lattice of $N$ sites. In \eqref{eq:General Evolution} we take Krauss operators of the form $A_{\nu}=m_{\nu}U_{\nu}$, where $U_{\nu}$
is an evolution under a non-interacting hamiltonian, and $m_{\nu}$ is a polynomial of order $r_{\nu}$ in fermion operators $a^{\dag},a$. 
The evolution under ${\cal L}$ of a general correlation function,
\begin{eqnarray}\label{k plus l corr}
\langle a_{i_{1}}^{\dag}..a_{i_{l_{1}}}^{\dag}a_{i_{(l_{1}+1)}}..a_{i_{(l_{1}+l_{2})}}\rangle\equiv \Tr\rho ~a_{i_{1}}^{\dag}..a_{i_{l_{1}}}^{\dag}a_{i_{(l_{1}+1)}}..a_{i_{(l_{1}+l_{2})}}
\end{eqnarray}
is given by
\begin{eqnarray}\label{eq:Hierarchy}
 & \langle a_{i_{1}}^{\dag}..a_{i_{l_{1}}}^{\dag}a_{i_{(l_{1}+1)}}..a_{i_{(l_{1}+l_{2})}}\rangle\longrightarrow\langle a_{i_{1}}^{\dag}..a_{i_{l_{1}}}^{\dag}a_{i_{(l_{1}+1)}}..a_{i_{(l_{1}+l_{2})}}\rangle+\\ \nonumber & \sum_{\nu}Tr~\rho U_{\nu}^{\dag}m_{\nu}^{\dag}[a_{i_{1}}^{\dag}..a_{i_{l_{1}}}^{\dag}a_{i_{(l_{1}+1)}}..a_{i_{(l_{1}+l_{2})}},m_{\nu}U_{\nu}]
\end{eqnarray}
where the normalization relation in \eqref{eq:General Evolution} was used.

The assumption that the $U_{\nu}$ are non interacting, means 
that $U_{\nu}^{\dag}a_{i}U_{\nu}=u_{\nu;ij}a_{j}$ for some unitary matrix $u_{\nu}\in U(N)$.
As a consequence the evolution of the $l_1+l_2$ correlation function \eqref{k plus l corr}, 
is related in (\ref{eq:Hierarchy}) to correlation functions of an order at most $l_1+l_2+2 \max_{\nu}(r_{\nu})$,
establishing a hierarchy of equations. 

We emphasize that the resulting state may be arbitrarily complex. Indeed, even when starting with a non-interacting thermal state, $\rho\sim exp(-h_{ij}a_{i}^{\dag}a_{j})$ and taking each $A_{\nu}$ a non interacting unitary, $\rho$ evolves into a sum of exponentials of fermion bi-linears. Such a state can be used to approximate any interacting state whose determinant quantum Monte Carlo description does not suffer from a sign problem \cite{grover2013quantum}.

Below, we list several fundamental operations under which the hierarchy {\it closes} at the two point
function level, for $G_{ij}\equiv\langle a_{i}^{\dagger}a_{j}\rangle$, inducing a map $G\to {\cal K}(G)$. We start with the obvious one:

(I) 
The non-interacting evolution ${\cal L}_{u}(\rho)=U\rho U^{\dagger}$, as described above, induces a map \[
G_{ij}  \rightarrow  {\cal K}_u(G)_{ij}\equiv(u^{\dagger}Gu)_{ij}
\]

We augment
the free evolution with the following types of operations acting on a single particle mode: particle detection, injection and extraction. Below, for simplicity of presentation we will associate the operation with the mode associated with site $i$.

Denote $P_{i}$ the matrix $(P_{i})_{ {m}  {n} }=\delta_{i {m} }\delta_{i {n} }$
the projection on site $i$, and $P_{i}^{\perp}=1-P_{i}$, we introduce:

(II) Particle detection at site $i$:
\[
{\cal L}_{D,i}(\rho)=n_{i}\rho n_{i}+(1-n_{i})\rho(1-n_{i})
\]
where $n_{i}=a_{i}^{\dag}a_{i}$. The induced map on $G$ is:
\[
{\cal K}_{D,i}(G)=P_{i}^{\perp}G P_{i}^{\perp}+P_{i}GP_{i}~.
\]
The process (II) may be viewed as a ``decoherence'' of
the correlations $G$ in between site $i$ and the rest of the lattice. As a linear super-operator on matrices, the measurement ${\cal K}_{D,i}$ has a simple spectrum. It acts as identity on matrices which do not mix site $i$ with the rest, hence the non-zero subspace of matrices has a dimension $1+(dim P^{\perp}_i)^2$. The complementary zero subspace is spanned by the off diagonal blocks, of dimensionality $2(dim P^{\perp}_i)$. 

(III) Removal of a particle from site $i$ is described by
\[
{\cal L}_{out,i}(\rho)=a_{i}\rho a_{i}^{\dagger}+(1-n_{i})\rho(1-n_{i})
\]
with the induced map on $G$:
\[
{\cal K}_{out,i}(G) =P_{i}^{\perp}G P_{i}^{\perp}~.
\]
As a super operator this simple map may be viewed as a projection on the space of matrices that do not have an $(i,j)$ or $(j,i)$ element for any $j$.

(IV) Finally, this operation injects a particle at site $i$:
\[
{\cal L}_{in,i}(\rho)=a_{i}^{\dagger}\rho a_{i}+n_{i}\rho n_{i}
\]
and induces the map 
\begin{equation}
{\cal K}_{in,i}(G)=P_{i}+P_{i}^{\perp}G P_{i}^{\perp}\label{eq:Injection}.
\end{equation}
We note that in contrast with $(I-III)$, the injection ${\cal K}_{in,i}$ is an in-homogenuous transformation on matrices, a property which we use below to compute steady states. 

It is also possible to add another two operations which correspond to "softer" particle motion into and out of the system, without performing a direct measurement on the system. These are described by:

($\tilde{\text{III}}$) Soft removal site $i$ is described by
\[
{\cal L}_{out,i,\epsilon}(\rho)=\epsilon(2-\epsilon) a_{i}\rho a_{i}^{\dagger}+(1-\epsilon n_{i})\rho(1-\epsilon n_{i})
\]
with the induced map on $G$:
\[ 
{\cal K}_{out,i,\epsilon}(G) =P_{i}^{\perp}G P_{i}^{\perp}+(1-\epsilon ) P_i G P_{i}^{\perp}+(1-\epsilon )P_{i}^{\perp} G P_i+(1-\epsilon
   )^2P_i G P_i.
\]
Here $0\leq \epsilon \leq 1$, with $\epsilon=1$ corresponding to the operation (III). Similarly, we have:

($\tilde{\text{IV}}$) Soft injection at site $i$:
\[
{\cal L}_{in,i,\epsilon}(\rho)=\epsilon(2-\epsilon)  a_{i}^{\dagger}\rho a_{i}+(1- \epsilon(1- n_{i}))\rho (1- \epsilon(1- n_{i}))
\]
and induces the map 
\begin{equation}
{\cal K}_{in,i,\epsilon}(G)=P_{i}^{\perp}G P_{i}^{\perp}+(1-\epsilon ) P_i G P_{i}^{\perp}+(1-\epsilon )P_{i}^{\perp} G P_i+(1-\epsilon
   )^2P_i G P_i+\epsilon(2-\epsilon)P_i \label{eq:SoftInjection}.
\end{equation}
Below, unless remarked differently, we will refer to both soft and hard process together, ommiting the $\tilde{}$ notation.
We can combine any of the site operations (II-IV) with the unitary evolutions (I) mixing the
the addressed site $i$ with the rest of the sites. When no particle injection is present, the particle extraction map will generically
 drive $G$ to $0$, i.e. $({\cal K}_{u}{\cal K}_{Out,i})^{n} \rightarrow 0$
\footnote{The limit $G \rightarrow 0$ is not at odds
with the validity of the map ${\cal L}$ at the level of density matrices: acting on density matrices ${\cal L}$
has to be positive and trace preserving. Here, the limit $G \rightarrow 0$ simply means $\rho\rightarrow |vac\rangle\langle vac|$ 
where the vacuum state $|vac\rangle$ is a perfectly normalizable state with
$G=0$.}. Similarly, adding particles by injection $({\cal K}_{u}{\cal K}_{In,i})^{n}$, with no extraction present, will result in $G_{ij} \rightarrow   \delta_{ij}$, when $n \rightarrow \infty$, which is the state where all sites are occupied. 

On the other hand the unitary evolution $(I)$ and the detection
process $(II)$ preserve the average particle number, i.e. $\langle \sum_i a^{\dag}_ia_i\rangle =\Tr~G$ remains constant under ${\cal K}_{u},{\cal K}_{M}$.

\subsection{Universality of the transformations (I,II,III,IV) on $G$}
The set of transformations (I,II,III,IV) generate all possible transformations on the two point function $G$, keeping $G$ a valid to point function by construction. In other words, given two valid correlation matrices $G_1$ and $G_2$, there is a set of operations of the form (I,II,III,IV) that will take us from $G_1$ to $G_2$.

{\it Proof:}
We have already seen that it is possible to get $G=0$ by emptying the system. It is therefore enough to show that we can get any $G$ starting from the zero matrix. 

To do so, let $u$ be unitary matrices that diagonalize  $G$ , i.e.:
\bea &
u^{\dag} G u={\cal K}_{u }(G)=diag(\lambda_{1},..,\lambda_{N})
\eea
Observing the operation \eqref{eq:SoftInjection}, and noting that $\lambda=(2-\epsilon)\epsilon$ for $\epsilon=1-\sqrt{\lambda}$ we have:
\bea
diag(\lambda_{1},0,..,0)={\cal K}_{in,1,\epsilon_1}(0)~~\epsilon_1=(1-\sqrt{\lambda_1})
\eea
similarly:
\bea
diag(\lambda_{1},\lambda_2,..,0)={\cal K}_{in,2,\epsilon_2}({\cal K}_{in,1,\epsilon_1}(0))~~\epsilon_2=(1-\sqrt{\lambda_2})
\eea
We can continue this way to populate the diagonal and get $diag(\lambda_{1},..,\lambda_{N})$. Finally, we undo the unitary $u$ and have
\bea
G={\cal K}_{u \dag}({\cal K}_{in,N,\epsilon_N}({\cal K}_{in,N-1,\epsilon_{N-1}}(....)))
\eea
with $\epsilon_i=(1-\sqrt{\lambda_i})$ at step $i$.

\subsection{Soft extraction by tunneling and removal from auxiliary site.}
We note that it is possible to induce the Kraus operators corresponding to the transformation
\[ \label{extract0}
{\cal L}_{out,0}(\rho)=\epsilon(2-\epsilon) a_{0}\rho a_{0}^{\dagger}+(1-\epsilon n_{0})\rho(1-\epsilon n_{0})
\]
with the induced map on $G$:
\[
{\cal K}_{out,0}(G) =P_{0}^{\perp}G P_{0}^{\perp}+(1-\epsilon ) P_0 G P_{0}^{\perp}+(1-\epsilon )P_{0}^{\perp} G P_0+(1-\epsilon
   )^2P_0 G P_0.
\]
without carrying out any direct measurement on the system, instead the measurements are carried out on outside the system. 
We can represent the operation of removing a particle from site $0$ by coupling the site by a tunneling Hamiltonian to an auxiliary site $e$, and making the "hard" removal on the site $e$. 

To make the derivation clear, let us denote by $\rho_S$ the density matrix of our system of $n$ fermionic sites. And the density matrix including the extra site $e$ is $\rho_{S+e}$.
We first perform operations on the larger system $\rho_{S+e}$, and compute the change in $\rho_S=\Tr_e \rho_{S+e} $ following the process.

The protocol is as follows.\\

(1) Site $e$ is decoupled from our system, and an operation of particle removal from $e$ is done. Thus
$$\rho_{S+e} \longrightarrow  (1-n_e )\rho_{S+e}
    (1-n_e )+a_e \rho_{S+e}  a_e^{\dagger }$$
This operation does not affect $\rho_S$. 
\\
(2) We apply the evolution with a tunneling between site $e$ and $0$, using the Hamiltonian $H_{t}\propto i(a_e^{\dagger
   }a_0-a_0^{\dagger }a_e )$. i.e. we evolve $\rho_{S+e}$ with:
\bea U_{\theta }=e^{\theta   (a_e^{\dagger
   }a_0-a_0^{\dagger }a_e )} \eea

Following these operations, we have to compute how $\rho_S$ transformed $\rho_S\rightarrow \rho_{S,1}\rightarrow \rho_{S,2}$. This can be done explicitly by choosing a basis for the Fock space. With Fermions we have to fix an ordering, and we take:
\bea
|m,\sigma ,\overset{\to }{k}\rangle =\text{  
   }\left(a_e^{\dagger }\right){}^m\text{ 
   }\left(a_0^{\dagger }\right){}^{\sigma
   }\left(a_1^{\dagger }\right){}^{k_1}
   \text{..}\left(a_{N-1}^{\dagger
   }\right){}^{k_{N-1}}\left(a_N^{\dagger
   }\right){}^{k_N}|\Omega \rangle
\eea
 where $m,\sigma,k_i\in\{0,1\}$.
 The reduced density matrix is computed as:
 \bea \label{reducedDen} &
 \langle \sigma ,\overset{\to }{k}|\rho_S|\sigma
   ',\overset{\to }{k'}\rangle =\langle 0,\sigma
   ,\overset{\to }{k}|\rho_{S+e}  |0,\sigma ',\overset{\to
   }{k'}\rangle +\langle 1,\sigma ,\overset{\to
   }{k}|\rho_{S+e}  |1,\sigma ',\overset{\to }{k'}\rangle
   = \\ \nonumber & \langle 0,\sigma ,\overset{\to
   }{k}|\rho_{S+e} +a_e \rho_{S+e}  a_e^{\dagger }|0,\sigma
   ',\overset{\to }{k'}\rangle 
 \eea

We now follow the steps outlined above.\\ 
(1) After step $1$, the total density matrix after particle removal from site $e$ is:
\bea
\rho_{S+e,1}=(1-n_e)\rho_{S+e}(1-n_e)+a_e \rho_{S+e}  a_e^{\dagger }
\eea
and the system density matrix is: $\rho_{S,1}=\rho_{S}$. We can also see this is to explicitly writing
\bea&
\langle \sigma ,\overset{\to }{k}|\rho_{S,1}|\sigma
   ',\overset{\to }{k'}\rangle =\langle 0,\sigma ,\overset{\to
   }{k}|\rho_{S+e,1} +a_e \rho_{S+e,1}  a_e^{\dagger }|0,\sigma
   ',\overset{\to }{k'}\rangle =\\ \nonumber & =\langle 0,\sigma ,\overset{\to
   }{k}|\rho_{S+e,1} |0,\sigma
   ',\overset{\to }{k'}\rangle = \langle \sigma ,\overset{\to }{k}|\rho_{S}|\sigma
   ',\overset{\to }{k'}\rangle  \Rightarrow \rho_{S,1}=\rho_{S}
\eea
\\
(2) We now apply the evolution $U_{\theta }$. We have 
$\rho_{S+e,2}=U_{\theta }\rho_{S+e,1}U_{\theta }^{\dagger } $, and therefore:
 \bea &
 \langle \sigma ,\overset{\to }{k}|\rho_{S,2}|\sigma
   ',\overset{\to }{k'}\rangle =\langle 0,\sigma ,\overset{\to }{k}|U_{\theta }\rho_{S+e,1}U_{\theta }^{\dagger } +a_e U_{\theta }\rho_{S+e,1}U_{\theta }^{\dagger } a_e^{\dagger }|0,\sigma
   ',\overset{\to }{k'}\rangle =\\ \nonumber & \langle
   0,0,\overset{\to }{k}|a_0^{\sigma }U_{\theta
   }\rho_{S+e,1}U_{\theta }^{\dagger } \left(a_0^{\dagger
   }\right){}^{\sigma '}+a_0^{\sigma }a_e U_{\theta
   }\rho_{S+e,1}U_{\theta }^{\dagger } a_e^{\dagger }
   \left(a_0^{\dagger }\right){}^{\sigma
   '}|0,0,\overset{\to }{k'}\rangle
 \eea
 To compute the matrix elements, we use the following properties of $U_\theta$:
 \bea
 U_{\theta }^{\dagger }|0,0,\overset{\to }{k'}\rangle
   =|0,0,\overset{\to }{k'}\rangle \text{           
   };\text{       }U_{\theta }^{\dagger
   }|1,1,\overset{\to }{k'}\rangle
   =|1,0,\overset{\to }{k'}\rangle
 \eea
 and the transformation:
 \bea &
 U_{\theta }^{\dagger }a_0U_{\theta }= \cos (\theta )
   a_0- \sin(\theta ) a_e\text{   }  \\ & U_{\theta }^{\dagger }a_eU_{\theta }= \cos (\theta )
   a_e+ \sin(\theta ) a_0\text{   }
 \eea
By commuting the $U_\theta$ operators through the $a_e,a_0$ operators we can now express the new matrix elements as function of $\theta$. We find that:
\bea &
 \langle \sigma ,\overset{\to }{k}|\rho_{S,2}|\sigma
   ',\overset{\to }{k'}\rangle =\delta _{\text{$\sigma $0}}\delta _{\sigma
   '0}\langle 0,0,\overset{\to }{k}|\rho_{S+e,1}
   |0,0,\overset{\to }{k'}\rangle
   + \sin^2(\theta )\delta _{\text{$\sigma
   $0}}\delta _{\sigma '0}\langle 0,1,\overset{\to
   }{k}|\rho_{S+e,1}|0,1,\overset{\to }{k'}\rangle
   +\\ \nonumber & \delta _{\text{$\sigma $1}}\delta _{\sigma
   '1}\cos^2(\theta )\langle 0,1,\overset{\to
   }{k}|\rho_{S+e,1}\text{  }|0,1,\overset{\to
   }{k'}\rangle +\\ \nonumber &  \cos (\theta ) \delta
   _{\text{$\sigma $1}}\delta _{\sigma '0}\langle
   0,1,\overset{\to }{k}|\rho_{S+e,1} |0,0,\overset{\to
   }{k'}\rangle + \cos (\theta ) \delta
   _{\text{$\sigma $0}}\delta _{\sigma '1}\langle
   0,0,\overset{\to }{k}|\rho_{S+e,1} |0,1,\overset{\to
   }{k'}\rangle \eea
We can identify the transformation on $\rho_S$ as:
   \bea&
   \rho _{S,2}=(1-n_0)\rho
   _{S}^{\theta
   }(1-n_0)+ \sin^2(\theta )
   a_0\rho _{S}^{\theta }a_0^{\dagger
   }+ \cos ^2(\theta )n_0\rho _{S} n_0+\\ \nonumber &  \cos (\theta )n_0\rho _{S} (1-n_0)+ \cos (\theta
   )(1-n_0)\rho _{S}^{\theta }n_0
   \eea
Rearranging the terms we finally have:
   \bea&
\rho _{S}\rightarrow   \rho _{S,2}=\left(1-\left(1- \cos (\theta
   )n_0\right)\right)\rho _{S} \left(1-\left(1- \cos (\theta
   )n_0\right)\right)+ \sin^2(\theta ) a_0\rho_{S} a_0^{\dagger }
   \eea
Identifying $\epsilon=1-\cos\theta$, and noting that $\sin^2{\theta}=\epsilon(2-\epsilon)$, we have recovered the map \eqref{extract0}.

\section{Non-Equilibrium Steady State Equation}
There are a myriad possible processes described by combinations of the operations $(I-IV)$. Here we concentrate on current generation processes as described by Eq.~\eqref{current injection process}, involves operations $I, {\text{III}}, {\text{IV}}$ resulting in the map:
\bea \label{G evolution} &
G\to  (1-r) u^{\dagger } G u+ r u^{\dagger }\{\alpha   ( (1-\epsilon_a
   P_a )G  (1-\epsilon_a P_a )+ (2\epsilon_a-\epsilon_a^2 )P_a ) +\\ \nonumber & 
   (1-\alpha ) ( (1-\epsilon_b P_b ) G (1-\epsilon_b P_b ) ) \}u .
  \eea 
This simple model allows for a substantial reduction of complexity from the full quantum problem of describing the evolution of $\rho$ into an evolution equation for the two point function $G_{ij}$, which can be tractable by either analytical or numerical methods. 
It is clear at this stage that we can access very interesting situations. 

To compute the eventual non-equilibrium steady state for \eqref{G evolution} it is convenient to view the transformation on $G$ from a point of view of
a super-operator. Here the $N\times N$ matrix $G$
is viewed as an $N^{2}$ dimensional vector, and the action of the evolution
${\cal L}$ on $\rho$ translates in \eqref{G evolution} into: 
\begin{equation}
G \rightarrow \Lambda G+g\label{eq:super operator},
\end{equation}
where $\Lambda$ is an $N^{2}\times N^{2}$ matrix, and $g$ is the inhomogeneous contribution due to the particle injection processes (\ref{eq:Injection}), and corresponding
to the term $ {r}   {\alpha}  (2\epsilon_a-\epsilon_a^2 ) u^{\dag}P_{a}u$ in \eqref{G evolution}. 

In general, whenever $g=0$, the long time behavior will be determined as usual
by the largest eigenvectors of $\Lambda$. However when $g\neq0$, the situation is somewhat different: Indeed, from
Eq. (\ref{eq:super operator}), we see that when $(1-\Lambda)$ is invertible, there exists a unique stationarity
$G$, that may be written in the form:
\[\label{super_steady_inv}
G_{steady}=(1-\Lambda)^{-1}g
\]
If $\Lambda-1$ is not invertible, i.e. there are steady states $\Lambda G_ {r} =G_ {r} $, it means that the evolution $u$ has an invariant subspace which does not include the sites $a,b$. In this case one has to work with a generalized inverse of $(\Lambda-1)$. A steady solution can either not-exist, or be non-unique of the form $G_{steady}\sim G_ {r} +(1-\Lambda)^{-1}g$. While inhomogenous equations are a common occurrence in the study of steady states in classical driven systems, they are used less in quantum processes, where evolution is unitary. A recent example of such a non-homogenous equation in a quantum context is the calculation of the expectation values of spin components in the steady state of a spin undergoing periodic laser pulses \cite{barnes2011electron,economou2014theory}.

We now apply these ideas to our current injection process described by \eqref{current injection process} and \eqref{G evolution}. Performing the inversion in superoperator space as in \eqref{super_steady_inv} in general is a daunting task. In the limit of $ {r}\ll1$, we were able to solve exactly for the degenerate perturbation theory to lowest order in $ {r} $, obtaining for the energy distribution $\Phi$ the result \eqref{energy distribution}. The derivation is somewhat lengthy and given in the next section. 

The $\cal{A},\cal{B}$ coefficients in \eqref{energy distribution} are given below. Define:
\bea\label{A and B coefficients}
& {\cal A}=\frac{2\alpha   (2-\epsilon_a )\text{  
   }(1-\alpha ) \epsilon_b^2 \epsilon_a
   Q_{\text{ab}}}{ ((2-\epsilon_a)
   (2-\epsilon_b )+2 Q_{\text{ab}}
   \epsilon_a \epsilon_b  (\alpha (2-
   \epsilon_a)+(1-\alpha )(2-\epsilon_b
   )))} \\ \nonumber & {\cal B}=\frac{\alpha  (2-\epsilon_a) \epsilon_a
   (2-\epsilon_b +2 \alpha  Q_{\text{ab}}
   \epsilon_a \epsilon_b )}{
   ((2-\epsilon_a) (2-\epsilon_b )+2 Q_{\text{ab}} \epsilon_a \epsilon_b 
   (\alpha (2- \epsilon_a)+(1-\alpha )(2-\epsilon_b
   )))}
\eea
where:
$$
\mu _k=2\left(\alpha \epsilon_a p_{a,k}+(1-\alpha) \epsilon_b
   p_{b,k}\right)~;~Q_{ab}=\Sigma_k\frac{p_{a,k}p_{b,k}}{\mu_k}
$$

We have verified the validity of the result numerically on numerous cases in addition to the one depicted in Fig. \ref{steadycurrent}(b).
We see that to leading order, $\Phi$ is independent of $r$. How can we understand this? Note that at $r=0$, there are infinitely many steady states (any $G$ such that $[G,h]=0$). However, when $r\neq 0$, $\Lambda$ stops being degenerate and it singles out a particular direction of breaking the degenerate space of matrices.

\subsection{Steady state distribution: Derivation}
Here we derive the formulas  \eqref{main_distribution},\eqref{A and B coefficients} for the non-equilibrium steady state energy distribution $\Phi$. 
We will study the steady state equation associated with the process \eqref{G evolution}, taking $\epsilon_a,\epsilon_b=1$ for simplicity, however the derivation with $\epsilon_a,\epsilon_b\neq 1$ follows along exactly the same lines. 
\bea & \label{steady state eqn}
{G_{steady}}=(1-r)\text{  }u^{\dagger } {G_{steady}} u+\\ & \nonumber  u^{\dagger }r \alpha  {}(P_a+P_{a{\perp}}{G_{steady}} P_{a{\perp}}{})u +  u^{\dagger }r (1-\alpha){}(P_{b{\perp}}{G_{steady}}
P_{b{\perp}}{})u
\eea
where $u=e^{-i \tau h_0}$. 

Below we label the eigenstates of $h_0$ by $n$, $h_0|n\rangle=E_n|n\rangle$, and would like to find the probability to find a state with energy $E_n$ occupied in the steady state. This probability is given by $\Phi_n\equiv Tr(\rho a^{\dag}_n a_n)=\langle n|G|n \rangle$.

For $r=0$, all states where $[G,h]=0$, are immediately invariant under time evolution. Therefore, in the limit of $r\ll 1$ we look for  an ansatz for the steady state ${G_{steady}}$ which is approximately diagonal. Let us write, in the energy basis, the ansatz: 
\bea & {G_{steady}}=\text{diag}(\{\Phi_1,...\}) +r D,\eea 
where ${\Phi_n}=\langle n|{G_{steady}}|n \rangle$ are the steady states occupations, and $D$ is an off-diagonal matrix in energy space.
Eq. \eqref{steady state eqn} becomes:
\bea &  
\Phi +r D=
(1-r) \Phi+(1-r) r u^{\dagger } D u+ r \alpha  u^{\dagger } P_au+\\ & \nonumber  r \alpha  u^{\dagger } {}(P_{a{\perp}}\Phi P_{a{\perp}}{})u+ 
u^{\dagger }r (1-\alpha){}(P_{b{\perp}}\Phi P_{b{\perp}}{})u+O(r^2)
\eea
We note that the zeroth order is eliminated and we wind up with:
\bea & D=- \Phi+ u^{\dagger } D u+    \alpha  u^{\dagger } P_a u+\alpha  u^{\dagger } {}(P_{a{\perp}} \Phi P_{a{\perp}}{})u+u^{\dagger } (1-\alpha)
{}(P_{b{\perp}}\Phi P_{b{\perp}}{})u\eea
Furthermore, note that both $D, u^{\dagger } D u$ are off-diagonal in energy. Therefore we have a closed equation for the diagonal elements:
\bea \label{diagonal equation} & 0=-{\Phi_n}+\alpha   p_{a,n} + \alpha  {  }{}(P_{a{\perp}}\Phi P_{a{\perp}}{})_{\text{nn}}+(1-\alpha) {}(P_{b{\perp}}\Phi
P_{b{\perp}}{})_{\text{nn}}.\eea
Explicitly,
\bea & {}(P_{a{\perp}} \Phi P_{a{\perp}}{})_{\text{nn}}=  {}(\Phi-P_a \Phi-\Phi P_a+P_a \Phi P_a{})_{\text{nn}}= {\Phi_n}-2 p_{a,n} {\Phi_n}+\Sigma
_lP_{a,\text{nl}}\Phi_lP_{a,\ln }\eea
where we have denoted $p_{a,n }=\langle n |P_a| n \rangle$ (and similarly $p_{b,n }=\langle n |P_b| n \rangle$) and $P_{a,ln }=\langle l |P_a| n \rangle$. Note that using $\Sigma _{l }P_{a,\text{nl}}P_{a,\ln }= p_{a,n} $ we can write  Eq. \eqref{diagonal equation}  as:
\bea & 0=\alpha   p_{a,n} -{\Phi_n}{}(\alpha  p_{a,n} +(1-\alpha)  p_{b,n} {}) 
+\Sigma _l{}({\Phi_l}-{\Phi_n}{}){}(\alpha  P_{a,\text{nl}}P_{a,\ln
}+(1-\alpha) P_{b,\text{nl}}P_{b,\ln }{}).\eea
At this point it is possible to argue that on the right,
$  | \Sigma _l{}({\Phi_l}-{\Phi_n}{}){}(\alpha  P_{a,\text{nl}}P_{a,\ln }+(1-\alpha) P_{b,\text{nl}}P_{b,\ln }{}) | $
is small, giving us a first guess for the answer:
\bea & {\Phi_n}\sim\frac{\alpha   p_{a,n} }{\alpha \text{  } p_{a,n} +(1-\alpha)  p_{b,n} }\label{naive}\eea
However, as we see below, it is possible to do better and solve equation \eqref{diagonal equation} exactly without this condition.
To do so notice that:
\bea & P_{a,\text{nl}}P_{a,\text{ln}}=|\langle n,a\rangle|^2 |\langle a,l\rangle|^2  \equiv  p_{a,n}  p_{a,l} \eea
Going back to  \eqref{diagonal equation} we write it as:
\bea & 0=\alpha    p_{a,n}   -2{}(\alpha  p_{a,n}   +(1-\alpha)  p_{b,n}   {}){\Phi_n}+  \Sigma_l{}(\alpha    p_{a,n}     p_{a,l}  +(1-\alpha)  p_{b,n}     p_{b,l}  {}){\Phi_l} \eea
We rewrite the equation as an in-homogenous linear equation:
\bea & {\cal Q}^2\overset{\to }{\Phi}=\alpha  Z_F {  }\overset{\to }{F}+V \overset{\to }{\Phi}.\eea
Here $\overset{\to }{F}$ is a unit vector defined by: 
\bea & \overset{\to }{F}=\frac{  p_{a,n}   }{Z_F}~~~;~~~ Z_F=\sqrt{\Sigma_n  p_{a,n} ^2 },\eea 
 ${\cal Q}$ is a diagonal matrix
\bea & {\cal Q}_{\text{nm}}=\delta _{\text{nm}}\sqrt{\mu_n}\text{    };\text{   }{\mu_n}= {2{}(\alpha \text{  }  p_{a,n}   +(1-\alpha) p_{b,n}  )}, \eea
and $V$ can be written in the form
\bea & V_{\text{nm}}= \alpha    p_{a,n}     p_{a,m}  +(1-\alpha) {  }  p_{b,n}    |
g_m | {}^2=  \alpha  Z_F^2 {  }|F\rangle \langle F|+(1-\alpha) Z_G^2 |G\rangle \langle G|.\eea 
The solution is given formally by:
\bea & \label{z expression 1} ({\cal Q}^2-V{})\overset{\to }{\Phi}=\alpha  Z_F {  }\overset{\to }{F} {  }\Longrightarrow  \overset{\to }{\Phi}=\frac{1}{{\cal Q}^2-V}\alpha {  }Z_F \overset{\to }{F}=\alpha  Z_F {\cal Q}^{-1}\frac{1}{1-{\cal Q}^{-1}V {\cal Q}^{-1}}{\cal Q}^{-1}\overset{\to }{F}.\eea
Next, we define the unit vector $  |F_Q\rangle$ as
\bea &  |F_Q\rangle=Z^{-1}_{\text{FQ}}{\cal Q}^{-1}|F\rangle,~~;~~Z_{\text{FQ}}^2=\Sigma  _n\frac{  p_{a,n} ^2 }{\mu_n Z_F^2}.\eea
Note the normalization $\|F_Q\|^2=1$.
Similarly we define
\bea &  |G_Q\rangle=Z^{-1}_{\text{GQ}}{\cal Q}^{-1}|G\rangle,~~;~~Z_{\text{GQ}}^2=\Sigma _n\frac{  p_{b,n} ^2 }{\mu_n Z_G^2}.\eea

Using these, \eqref{z expression 1} is expressed as:
\bea \label{z expression 2} & \overset{\to }{\Phi}=  \alpha  {\cal Q}^{-1}{}(\frac{Z_FZ_{\text{FQ}}}{1-\alpha  Z_F^2 Z_{\text{FQ}}^2 |F_Q\rangle \langle F_Q|-(1-\alpha) Z_G^2 Z_{\text{GQ}}^2
|G_Q\rangle \langle G_Q| {    }} {})|F_Q\rangle\eea

In the next step we use the following relation:
\bea \label{Relation2} & \frac{1}{1+a |v\rangle \langle v|+b |u\rangle \langle u|}|v\rangle = \frac{1}{1+a+b+a b{}(1-{}|\langle v,u\rangle| ^2{})}{}\{(1+b)|v\rangle -b
\langle u,v\rangle |u\rangle {}\},\eea
which holds for normalized vectors $||u||=||v||=1$. We are not aware if the expression \eqref{Relation2} appears in the literature, but it can be verified explicitly by multiplying both sides by $(1+a |v\rangle \langle v|+b |u\rangle \langle u|)$.

We will use \eqref{Relation2} on  \eqref{z expression 2}, with $|F_Q{}\rangle ,|G_Q{}\rangle $ playing the role of $|u\rangle,|v\rangle$. Thus, we take in  \eqref{Relation2}:
\bea
a\rightarrow  -\alpha  Z_F^2 Z_{\text{FQ}}^2    ~~~~;~~~~ b\rightarrow   -(1-\alpha) Z_G^2 Z_{\text{GQ}}^2, 
\eea
and 
\bea & c\equiv{}\langle F_Q|G_Q{}\rangle =\Sigma _n\frac{1}{Z_GZ_FZ_{\text{FQ}}Z_{\text{GQ}}}\frac{  p_{b,n}     p_{a,n}  }{\mu_n}= \frac{1}{Z_GZ_FZ_{\text{FQ}}Z_{\text{GQ}}}\Sigma _n\frac{  p_{b,n}     p_{a,n}   }{\mu_n}\eea
noting
\bea & Z_FZ_{\text{FQ}}=\sqrt{\Sigma _n\frac{  p_{a,n} ^2 }{\mu_n}}\text{    };\text{   }Z_GZ_{\text{GQ}}=\sqrt{\Sigma _n\frac{ p_{b,n} ^2}{\mu_n}}\eea
we have
\bea & c=\frac{1}{\sqrt{(\Sigma
_l\frac{  p_{a,l} ^2 }{\mu_l})(\Sigma _l\frac{  p_{b,l} ^2 }{\mu_l})}}\Sigma _n\frac{  p_{b,n}    p_{a,n}   }{\mu_n}\eea
Using these expressions with \eqref{Relation2} and \eqref{z expression 2} we find:
\bea & \nonumber  {\Phi_n}= \frac{\alpha  Z_FZ_{\text{FQ}}}{\sqrt{\mu_n}}{}({}(\frac{1}{1-\alpha  Z_F^2 Z_{\text{FQ}}^2 |F_Q\rangle \langle F_Q|-(1-\alpha) Z_G^2 Z_{\text{GQ}}^2
|G_Q\rangle \langle G_Q| {    }} {})|F_Q\rangle {}){}_n=\\ & \nonumber  \frac{\alpha  Z_FZ_{\text{FQ}}}{\sqrt{\mu_n}} \frac{1}{1-\alpha  Z_F^2
   Z_{\text{FQ}}^2-(1-\alpha)  Z_G^2 Z_{\text{GQ}}^2+\alpha  Z_F^2
   Z_{\text{FQ}}^2(1-\alpha)  Z_G^2 Z_{\text{GQ}}^2 (1- |
   c| ^2 )}\times \\ & \nonumber \{(1-(1-\alpha)  Z_G^2
   Z_{\text{GQ}}^2 ) \langle n|F_Q\rangle +(1-\alpha)  Z_G^2
   Z_{\text{GQ}}^2 c^*  \langle n|G_Q\rangle \} =\\ & \nonumber
\frac{\alpha  }{\mu_n}  \frac{(1-(1-\alpha)  Z_G^2
   Z_{\text{GQ}}^2 )  p_{a,n}  +(1-\alpha) 
   Z_FZ_{\text{FQ}} Z_G Z_{\text{GQ}} c^*  p_{b,n}  }{1-\alpha  Z_F^2
   Z_{\text{FQ}}^2-(1-\alpha)  Z_G^2 Z_{\text{GQ}}^2+\alpha  Z_F^2
   Z_{\text{FQ}}^2(1-\alpha) Z_G^2 Z_{\text{GQ}}^2 (1- |
   c | ^2 )}. \eea
   
Denoting
\bea
Q_{ \text{aa}}=Z_F^2
   Z_{\text{FQ}}^2=\Sigma
_l\frac{  p_{a,l} ^2 }{\mu _l}\text{  };\text{   }Q_{ \text{bb}}=Z_G^2
   Z_{\text{GQ}}^2=\Sigma _l\frac{  p_{b,l} ^2 }{\mu _l}\text{
  };\text{    }Q_{ \text{ba}}=\Sigma _l\frac{ p_{a,l}  p_{a,l}  }{\mu _l}, \eea
we find that:
\bea \label{phi_before_simp}  & {\Phi_n}= \frac{\alpha }{ \mu _n}\frac{ {}(1-(1-\alpha) Q_{ \text{bb}} {})  p_{a,n}   +(1-\alpha) Q_{ \text{ba}}  p_{b,n}  }{1-\alpha  Q_{ \text{aa}}-(1-\alpha) {  }Q_{ \text{bb}}+\alpha  {  }(1-\alpha) {}( Q_{ \text{aa}}Q_{ \text{bb}}-Q_{ \text{ba}}{}^2{})}.\eea
As a final simplification we note that:
\bea
2 \alpha  Q_{\text{ba}}+2(1-\alpha)  Q_{\text{bb}}=\Sigma _l \frac{ p_{b,l}   (2 \alpha  p_{a,l} +2(1-\alpha) p_{b,l}  )}{2\left(\alpha \text{  } p_{a,l} +(1-\alpha)  p_{b,l} \right)}=\Sigma_l p_{b,l} =1, 
\eea
and similarly we have: $2 \alpha  Q_{\text{ba}}+2(1-\alpha)  Q_{\text{bb}}=1$. Using these relations in Eq. \eqref{phi_before_simp}, we can express the final result in terms of $Q_{\text{ab}}$ alone, finding:
\bea
\Phi _k=\frac{{\cal{A}}+{\cal{B}}\frac{ p_{a,k}}{p_{b,k}}}{(1-\alpha) +\alpha  \frac{
   p_{a,k}}{p_{b,k}}}\text{      };\text{    }{\cal{A}}=\frac{2 (1-\alpha ) \alpha 
   Q_{\text{ba}}}{1+2 Q_{\text{ba}}}\text{   },\text{  }{\cal{B}}=\frac{\alpha  \left(1+2
   \alpha  Q_{\text{ba}}\right)}{1+2 Q_{\text{ba}}}.
\eea
as mentioned, the derivation with $\epsilon_a,\epsilon_b \neq 1$ follows exactly the same line, giving the coefficients \eqref{A and B coefficients}.

\subsection{Kossakowski-Lindblad limit}
In the Kossakowski-Lindblad limit, the treatment is considerably simpler. Starting with:
\bea\nonumber
\dot{\rho }=\frac{1}{i \hbar }[H,\rho ]+ \gamma _{a}
   (a_a^{\dagger }\rho  a_a\!-\!\frac{ \{\rho,a_a a_a^{\dagger } \}}{2} )+\gamma_{b} (a_b\rho  a_b^{\dagger}\!-\!\frac{\{\rho ,
   a_b^{\dagger }a_b \} }{2} )
\eea
the equation for $G$ is:
\bea
\dot{G}=-\frac{i}{\hbar } [G,h^t]-
   \{\frac{\gamma _a P_a+\gamma
   _b P_b}{2},G\}+ \gamma_a P_a.
\eea
The steady state obeys $\dot{G_{steady}}=0$, we again set $G_{steady}=\Phi+r D$  where $\Phi$ is diagonal and $D$ is strictly off diagonal in energy, and assume that $r\rightarrow 0$ when $\gamma_a$ and $\gamma_b$ are approaching zero. We take a diagonal matrix element of the equation to find, in lowest order in $r$ that
\bea
\Phi _k \left(\gamma _a P_a+\gamma
   _b P_b\right)_{\text{kk}}=\gamma
   _a \left(P_a\right)_{\text{kk}}\text{   }\Rightarrow~~
\Phi_k\equiv \frac{\gamma _a p_{a,k}}{\gamma _ap_{a,k}+\gamma _bp_{b,k}}.
\eea
Setting $\gamma_a=r\alpha \epsilon_a$, $\gamma_b=r(1-\alpha) \epsilon_b$, as representing the appropriate rates in the process described in \eqref{current injection process} we recover \eqref{main_distribution}. 

\begin{figure} \centering 
\includegraphics[width = 0.6\textwidth]{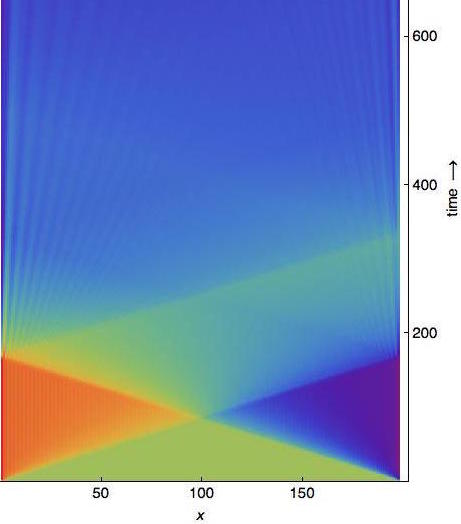} 
 \includegraphics[width = 0.6\textwidth]{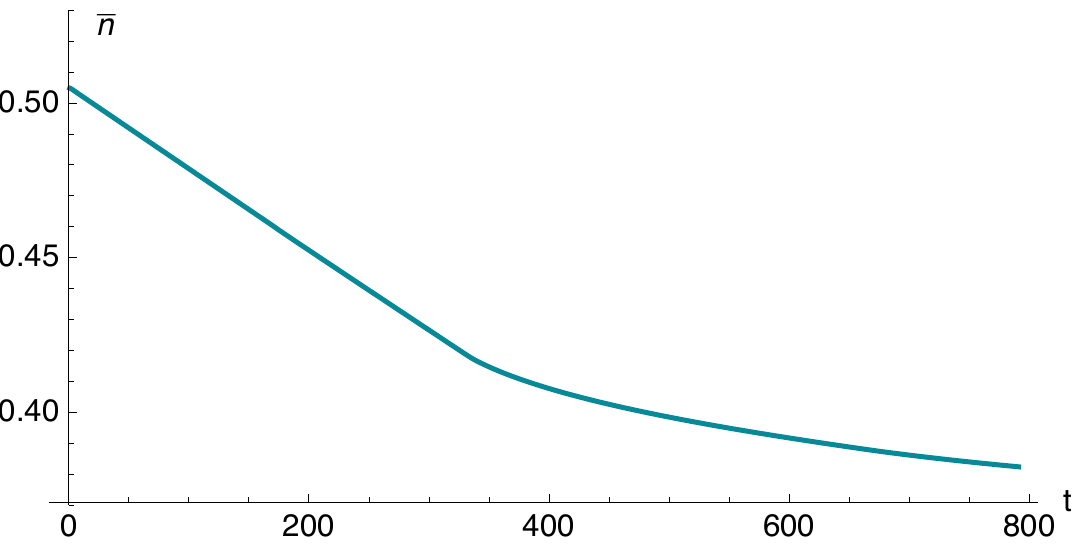}\caption{ Density depletion in a system where particles are extracted from the right at higher rate than injected on the left, here $r=1$ and $\alpha=0.3$, initial state is the half filled ground state of $H_{hop}$. Left: Real space evolution of local density. Right: Evolution of space averaged density. }\label{imbalanced}
\end{figure}
\begin{figure} \centering 
\includegraphics[width = 0.6\textwidth]{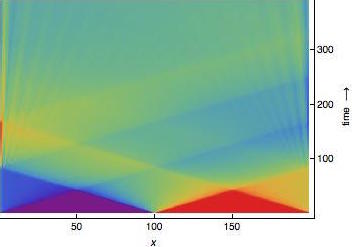} 
\includegraphics[width = 0.6\textwidth]{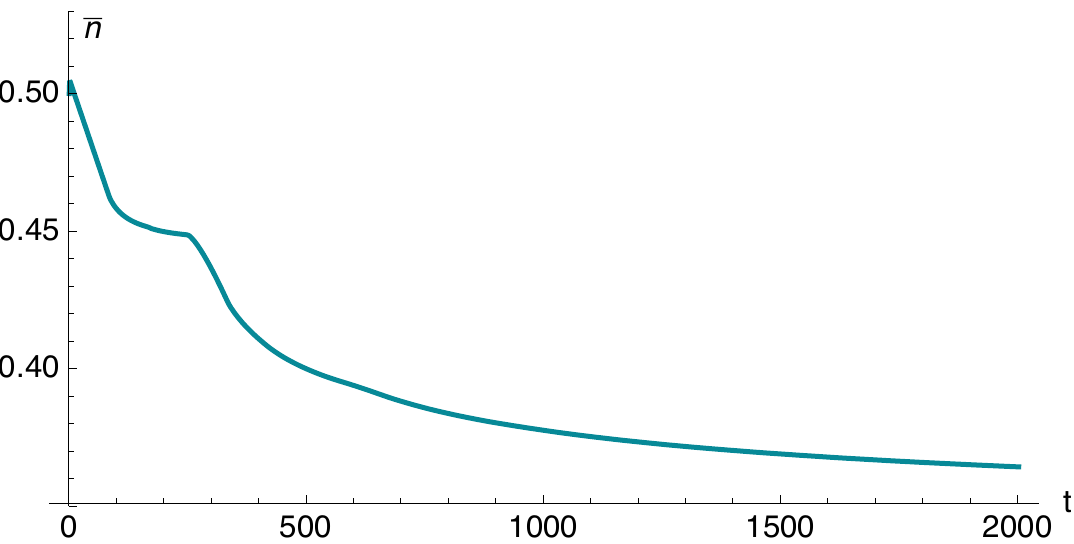}\caption{Density evolution under the same dynamics as Fig. \ref{imbalanced}, however with a domain wall as an initial state. Here the depletion happens in two steps, but eventually reaches the same asymptotic value ($\bar{n}\sim0.38$) as in Fig. \ref{imbalanced}.}\label{domaindep}
\end{figure}

\section{Examples of dynamics and dependence on initial condition}
The dependence of the dynamics on the initial condition is of interest by itself. While in Fig. \ref{steadycurrent}, we started the evolution from the vacuum state, 
in Fig. \ref{imbalanced}, we describe such a process where the system is started off as the ground state of $H_{hop}$.
The evolution happens in stages. In the initial stage of evolution we observe two shock wave fronts: one propagating with a region of reduced density from the right, collides with a front of enhanced density propagated from the left. It is interesting to note that the evolution is on a faster time scale than the speed of propagation of a wave-packet localized at a point by free evolution. In the context of classical non equilibrium processes, shock waves have been described for the asymmetric exclusion process in e.g.\cite{kolomeisky1998phase} (It is possible to use the present system also to describe such situations, however this will be done elsewhere).

As the fronts collide the imbalance between the left and right sides of the chain starts to decrease. 
Finally, soliton like density packets of different velocities, are observed at longer time scales, and may be related to the soliton described in \cite{bettelheim2006orthogonality} in the context of the orthogonality catastrophe. It is interesting to note the injected particles traveling from the left travel with faster velocities compared to their partners from the other side. 

In Fig. \ref{imbalanced} we show the average particle density $\bar{n}\equiv N^{-1}Tr G$.
One of the interesting features observed is a qualitative change in the slope of  $\bar{n}(t)$ around $350$ iterations. This change seems to correspond to the annihilation of the high density front coming from the left. To check this behavior, we consider, in Fig. \ref{domaindep}  the evolution when the initial stage is asymmetric itself: Here in the initial stage all sites $i$ on the left, $i<100$, are empty, while all sites on the right $i>100$ are occupied. This state evolves through four fronts that collide and eventually annihilate. Note that for coherent evolution from such an initial state, it has been shown that the front propagation has a scaling $1/t^{3}$ \cite{eisler2013full}. 
In the context of evolution of magnetization in a spin chain the evolution of initial domain wall was studied in \cite{antal2008logarithmic}.

Comparing the density evolution in Fig. \ref{imbalanced} and Fig. \ref{domaindep}, we see that there is a transient behavior associated with the different nature of the initial states, and their stages of evolution. In Fig. \ref{domaindep}, there is a noticeable change in depletion rate around $100$ and $300$ iterations, the first kink corresponds the initial high density region on the right hitting the left side: at that point injection of particles becomes harder for a while and $|\partial_t\bar{n}|$ decreases until the density goes down enough on the left. The second kink is observed when the high density region is reflected back to the right: extracting particles on the right is then easier and $|\partial_t\bar{n}|$ grows.
At long times the density seems to decay asymptotically as $1/t$ towards the non-equilibrium steady state density.

\section{Summary}
We presented a class of non equilibrium quantum processes that correspond to closed hierarchies of evolution equations, and can thus be studied numerically efficiently. We have used this idea to explore non-equllibrium generation of currents and approach to steady states. We remark that the resulting states may also be viewed as Floquet states, and we have thus supplied a particular way of engineering such states, that may be of interest in the context of topological Floquet states\cite{kitagawa2010topological,lindner2011floquet,gu2011floquet} and generation of topological states via dissipation \cite{diehl2011topology}.  Moreover, the energy distribution $\Phi_k$ should be studied further: one can hope to test the resulting highly excited current carrying steady states in a variety of settings from cold atoms to mesoscopic systems and spin chains. We emphasize that our result does not rely on integrability in the sense of Bethe Anzats that is useful in one dimension and has been used in studies of dissipative spin chains. Thus, our treatmentis available for periodically driven fermion systems that do not correspond to spin chains, and most importantly, to higher dimensional systems.

\textbf{Acknowledgement} It is a pleasure to thank E. Altman, T. Hughes, E. Kolomeisky and KW Kim for insightful discussions, as well as useful suggestions by J. Avron and L. Vidmar.
The work was supported by the NSF CAREER grant DMR-0956053.


\end{document}